\documentstyle[prl,aps,epsfig,floats,preprint,amssymb]{revtex}
\tightenlines
\begin{document}
\draft
\title{Objective Bayesian statistics}
\author{0.-A. Al-Hujaj and H.L. Harney}
\date{\today}
\address{MPI f\"ur Kernphysik,  Postfach 103980, 69029 Heidelberg, Germany}
\maketitle
\begin{abstract}
  Bayesian inference~--- although becoming popular in physics and
  chemistry~--- is hampered up to now by the vagueness of its
  notion of prior probability. Some of its supporters argue that
  this vagueness is the unavoidable consequence of the
  subjectivity of judgements~--- even scientific ones.
  We argue that priors can be defined uniquely if the
  statistical model at hand posses a symmetry and if the ensuing
  confidence intervals are subjected to a frequentist
  criterion.
  Moreover, it is shown via an example taken from recent experimental
  nuclear physics, 
  that this procedure can be extended to models
  with  broken symmetry.
\end{abstract}

\pacs{2.50 Kd, 2.50 Wp, 6.20 Dk}
\narrowtext

\section*{Introduction}

Bayesian inference is becoming popular in the physical sciences. It
finds eloquent and noteworthy defenders
\cite{Anderson:1992,Dagostini:1995}, it is treated in a growing number
of
textbooks\cite{Berger:1985,Lee:1989,Press:1989,Howson:1993,Bernado:1993},
an annual series of conferences\cite{Maximum} as well as numerous
articles report on its applications. Even the ``Guide to the
expression of uncertainty in measurement'' \cite{ISO:92} supported by
the International Organization for Standardization (ISO), the Bureau
International des Poids et M\'{e}sures (BIPM) and other international
organizations implicitly favors Bayesian statistics~--- as D`Agostini
points out (in Secs. 2.2 and 14.2 of \cite{Dagostini:1995}). Some
authors that have collaborated in the formulation of the Guide
\cite{ISO:92} and of the German standard \cite{DIN:85}, adhere in
their publications to ``a Bayesian theory of measurement uncertainty''
\cite{Weise:1992}. The change of paradigm from frequentist to Bayesian
statistics (for this antagonism see e.g. the introduction of
\cite{Weise:1992} and ch. 7 of \cite{Barlow:1989}) is taking 
place despite the fact, that the
disturbing vagueness of Bayesian prior probabilities persists up to
now. D`Agostini \cite{Dagostini:1995} eloquently holds that the prior
probabilities are the mathematical representation of an unavoidable
subjectivity of judgments~--- even of scientific ones.  In our opinion,
however,  a criterion exists that prior probabilities should meet~---
at least approximately~--- and which in many cases does not leave any
freedom in their choice. This criterion is a consequence of the fact
that Bayesian inference has a frequentist interpretation~--- as will be
explained below. This fact has been pointed out earlier
(by Welch, Peers, Stein and
Villegas~\cite{Welch:1963,Stein:1965,Villegas:1977b,%
Villegas:1981,Villegas:1990}) in the literature of mathematical
statistics but it seems rarely known in practice. In the present
note, the criterion is described as well as the circumstances
under which it is met. They amount to the existence of a
symmetry of the model that states the relation between event and
hypothesis. The Bayesian prior is then a Haar measure of the
symmetry group. By helps of a realistic example taken from
current nuclear physics it is shown that this procedure can be
extended to the usual case of models with broken
symmetry. Finally the same example shows that the popular likelihood method~---
as described in \cite{Eadie:1971}--- yields confidence intervals of
lower reliability than the Bayesian  procedure.

\section*{Bayesian Statistics}

Bayesian statistics is a very useful tool for
statistical inference.
Let  $x$ denote the continuous event and $\xi$ the continuous hypotheses.
 Suppose that the conditional probability $p(x|\xi)dx$~--- the
 above mentioned model~--- as well as the event $x$ are given.
 The problem of statistical inference
is: What can we learn from $x$ about $\xi$? The Bayesian
answer \cite{Bayes:1763} is a conditional probability $P$ for $\xi$
given $x$ which can be expressed in terms of the given
distribution p,
\begin{equation}
  \label{Bayesinverse}
  P(\xi|x)d\xi=\frac{p(x|\xi)\mu(\xi)}{\int d\xi'\,p(x|\xi')\mu(\xi')}d\xi.
\end{equation}
Here, $\mu(\xi)$ is the prior distribution assigned to $\xi$ in
the absence of or ``prior to'' any experimental evidence. Hence,
Bayesian statistics relies on the assumption that $\mu(\xi)$ is
a meaningful object.

The choice of the prior is the problem of Bayesian theory,
because it is not clear what we can know ``prior'' about $\xi$.

\section*{Subjective Interpretation}

The view held e.g. in the recent article by D'Agostini \cite{Dagostini:1995}
 or the recent textbook by 
Howson and Urbach \cite{Howson:1993} is that the choice of
$\mu(\xi)$ should be left to the good taste and the experience of
the scientist analyzing the event $x$. They argue that a such a
subjective element should be in any honest theory of inference
since it makes explicit the subjectivity of any judgement~---
including the scientific ones: nothing can be known about
$\xi$ in an objective way prior to the event. 
In subjective interpretation Bayesian probabilities reflect 
personal confidence in hypotheses. It can be expressed in bets
(see \cite{Howson:1993}).

 This view does in principle not
preclude objectivity altogether: Objectivity exists as the
limiting consequence of an infinite number of recorded
events. Indeed one can show that in this limit Bayesian inference
becomes independent of $\mu(\xi)$ in the sense that $p(x|\xi)$
then tends towards a $\delta$-distribution with respect to
$\xi$~--- centered at the true value $\hat{\xi}$ of the hypothesis.

\section*{Objective Bayesian statistics}

The subjective interpretation~--- taken literally and without
appealing to some common sense~--- allows anything. This is
obvious from eq.~(\ref{Bayesinverse}): if one is free to choose
$\mu(\xi)$, one can generate any $P(\xi|x)$~--- given  a finite
number of events. To avoid this, we 
prefer to look for some criterion that would severely restrict
the class of allowed priors. Fortunately there is a very natural one.

Consider a Bayesian confidence area ${\cal
  A}(x,C)$. It shall satisfy
\begin{equation}
  \int\limits_{{\cal A}(x,C)}d\xi\,P(\xi|x)\equiv C,
\end{equation}
which is usually stated in the form: ``given the event $x$, the
hypothesis lies with confidence $C$ in the area ${\cal A}$''. We
want to reformulate this in a way which turns the vague notion of
confidence into probability and by the same token defines the
desired criterion.

Imagine an ensemble $X$ of events $x$ with relative frequencies
$p(x|\hat{\xi})dx$. Let $x$ run over the ensemble and suppose
that from every $x$ the confidence area ${\cal A}(x,C)$ is derived
in a unique way, e.g. by determing the smallest one. This yields
an ensemble of confidence areas. The criterion then is: The prior
must lead to an ensemble of confidence areas such that they cover
the true value $\hat{\xi}$ with probability $C$.

Since this gedanken experiment~--- which can even be realized via
Monte Carlo simulation~--- equates $C$ with a well defined
frequency (to cover $\hat{\xi}$), the criterion turns confidence
into frequentist probability. In short: We require that
Bayesian confidence areas are frequentist confidence areas. It
is proven in the mathematical literature \cite{Stein:1965} that
this criterion can be met exactly: Let the conditional probability 
$p(x|\xi)dx$ be invariant under a Lie group $\mathbb{G}$
represented as transformations of $x$ and $\xi$, i.e.
\begin{equation}
  \label{invariance}
  p(x|\xi)dx=p(G_\rho x|G_\rho \xi)d G_\rho x, \qquad G_\rho\in\mathbb{G},
\end{equation}
suppose furthermore that the definition of the confidence area ${\cal
  A}$ is invariant under $\mathbb{G}$, i.e
\begin{equation}
  \label{area}
  {\cal A}(G_\rho x,C)=G_\rho {\cal A}(x,C),
\end{equation}
then the \emph{right Haar measure} of the symmetry group is a
suitable prior in the sense of the criterion.
It is necessary for this, that the hypothesis can be
identified with the symmetry group of the conditional
probability, i.e. for every two hypotheses
$\xi_1$,$\xi_2$  there must be exactly one transformation
$G_{\rho}\in \mathbb{G}$ such that $\xi_1=G_\rho \xi_2$. 
Then the uniqueness of the right Haar measure implies the
uniqueness of the prior and one has
\begin{equation}
  \label{haarmeasure}
  \mu(\xi)=\left[\left.\frac{\partial
    G_\rho\xi}{\partial\rho}\right|_{\rho=0}\right]^{-1},
\end{equation}
which is the inverse Jacobian of the transformation $G_\rho$
taken at the unit element of the group.

Note that the very interesting case of ${\cal A}$ to be the smallest
area of confidence $C$, satisfies eq. (\ref{area}), if the volume of
${\cal A}$ is determined by help of the measure that is invariant
under $\mathbb{G}$, i.e. the left Haar measure.

The prior distribution is not independent of
the statistical model; rather it is defined by the structure of
the model (see \cite{Villegas:1981,Villegas:1977a}). 
It is prior to the observations. Since the above
symmetry uniquely defines the prior and since it ensures a
frequentist interpretation of Bayesian confidence intervals and
since frequentist probability is often termed ``objective''
probability, one may call the present procedure \emph{objective}
Bayesian statistics. However, we return to the word ``Bayesian''
implying for the rest of the paper the qualification
``objective''. By the following example, we show that this
concept can be extended to models that lack the symmetry (\ref{invariance}).

\section*{Example}

The measurement problems encountered in science, often
do not have a symmetry (\ref{invariance}). However,  Bayesian
inference based on a multi-dimensional event can usually be broken
down into a succession of Bayesian arguments on more elementary events
until a starting point that possesses a symmetry has been found. To be
definite let us consider the parity violating matrix elements measured
by the TRIPLE collaboration in resonant p-wave scattering of polarized
neutrons on heavy nuclei
\cite{Bowman:1990,Frankle:1992,Zhu:1992,Bowman:1995}.  For the present
purpose it is not necessary to understand  the
details of those experiments. It suffices to know that the
 parity violating matrix elements~--- the
events~--- have the Gaussian probability
\begin{equation}
  \label{gaussprob}
  g(x|\sqrt{M^2+\varepsilon^2})dx=\frac{dx}{\sqrt{2 \pi
      (M^2+\varepsilon^2)}}\exp\left(
  -\frac{1}{2}\frac{x^2}{M^2+\varepsilon^2}\right)
\end{equation}
where $\varepsilon$ is an experimental error~--- supposed to be
given~--- and $M$ is the root mean square parity violating matrix
element. The latter one is the ``hypothesis'' to be determined.
There is no symmetry  (\ref{invariance}) relating $x$ and
$M$. However, the probability of eq. (\ref{gaussprob}) remains
invariant under simultaneous change of the scale of $x$ and
  $\xi=\sqrt{M^2+\varepsilon^2}$. 
Hence, Bayesian inference should be done with respect to $\xi$
rather than
 $M$. Afterwards~--- when the posterior distribution of
$\xi$ has been constructed~--- one can derive statements about
$M$. E.g. one can decide whether with
sufficiently high confidence $\xi$ is larger than $\varepsilon$ and
thus $M>0$. The symmetry considerations require
\begin{equation}\label{prior}
\mu(\xi)=\frac{1}{\xi}
\end{equation}
as prior, which is the Haar measure of the Lie group of scale
changes.

The problem of \cite{Bowman:1990,Frankle:1992,Zhu:1992,Bowman:1995} is
complicated by the fact, that one usually does not know the total
angular momentum of the p-wave
resonance. Only p$_{1/2}$ resonances can show parity violation.
 If the event $x$ is gathered in a p$_{3/2}$ resonance   nothing can
 be learned about $M$ and the
distribution of the event is $g(x|\varepsilon)$. One knows, however,
the probability $q_p$ of the occurrence of p$_{1/2}$ resonances. Again it
is not necessary to discuss here the contents of nuclear physics that
create this complication. It suffices to state, that the event $x$
follows with probability $q_p$ the distribution $g(x|\xi)$ and with
probability $1-q_p$ the distribution $g(x|\varepsilon)$, so that one
has
\begin{displaymath}
  p(x|\xi)=q_p g(x|\xi)+(1-q_p)g(x|\varepsilon).
\end{displaymath}
The presence of the second term on the right hand side precludes any
symmetry  (\ref{invariance}) 
relating $x$ and $\xi$~--- except in the limit
of $\varepsilon\rightarrow 0$ when $g(x|\varepsilon)\rightarrow
\delta(x)$ and $p(x|\xi)$ recovers the symmetry under scale changes.
The full experiment of
\cite{Bowman:1990,Frankle:1992,Zhu:1992,Bowman:1995} probes $n$
resonances $i=1,\ldots,n$ under varying experimental conditions, so
that $\varepsilon=\varepsilon_i$ becomes a function of the resonance
$i$.  This alone precludes any symmetry of the multidimensional
problem.

If, however, one knows for one of the resonances that it is a p$_{1/2}$
case, say for $i=1$, then the event measured there has the
distribution $g(x_1|\xi)$; the symmetry is scale-invariance and
the prior is (\ref{prior}); one  can use the
event $x_1$  to construct the Bayesian inverse
$P_1(\xi|x_1)$; it can be injected as prior distribution into the analysis of
the results at $i=2,\ldots,n$. This procedure  leads to a
posterior distribution $P(\xi|x_1,\ldots,x_n)$. In this way, one finds
a starting point of the whole analysis which has a well defined
symmetry and therefore a well defined prior. Let us call the whole
procedure an approximate Bayesian (AB) one.

\begin{figure}[htbp]
  \epsfig{file= 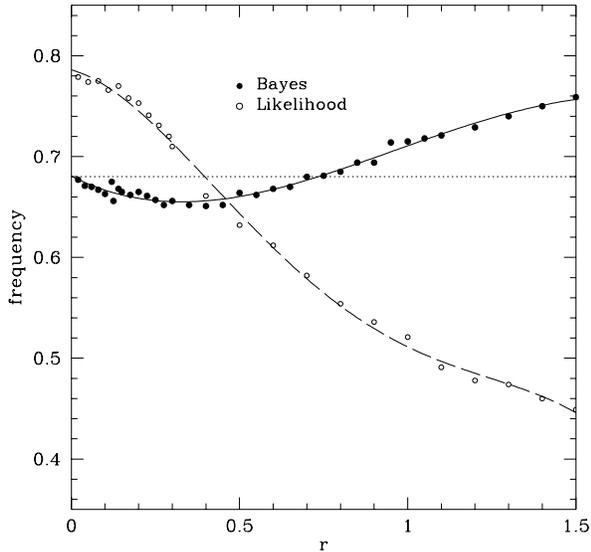,height=8cm,width=8cm}
  \caption{Bayesian inference is superior to the likelihood
    method: The frequency of the true value of the parity violating
    matrix element to lie in the 68\%-confidence interval is plotted
    against the size of the experimental error. Details are explained
    in the text.}
  \label{fig:Bayes-Likelihood}
\end{figure}

It is not our purpose to reanalyze the data of refs.
\cite{Bowman:1990,Frankle:1992,Zhu:1992,Bowman:1995}. We only want to
demonstrate, that the AB analysis  satisfies the criterion
described above to a good approximation. For the purpose of this, the
AB  analysis has 
been subjected to a Monte Carlo test with parameters close to those
of the experimental cases
\cite{Bowman:1990,Frankle:1992,Zhu:1992,Bowman:1995}. The number of
resonances was chosen to be $n=15$. The errors $\varepsilon_i$,
$i=1,\ldots,15$ have been drawn from an exponential distribution with
mean value $r\hat{M}$. This allows one to study (on
fig. \ref{fig:Bayes-Likelihood}) 
the result as a function of $r=\bar{\varepsilon}/\hat{M}$~--- the mean
error  relative to the true value $\hat{M}$ of the r.m.s. parity
violating matrix element. 
In the experiments
\cite{Bowman:1990,Frankle:1992,Zhu:1992,Bowman:1995}, $r$ ranged from
0.23 up to the order of unity.
 The coordinates $x_i$  of the event  $x=(x_1,\ldots,x_{15})$
were generated in two steps. First one decides  with
probability $q_p=\frac{1}{3}$, whether the quantity $x_i$ should belong to the
p$_{1/2}$-wave resonances. If yes then $x_i$ was drawn from an
ensemble with distribution $g(x_i|\sqrt{\hat{M}^2+\varepsilon_i^2})$, else the
distribution $g(x_i|\varepsilon_i)$ was applied. The vector $x$~--- without
the information from which of the two ensembles anyone of the $x_i$
comes~--- is  equivalent to the experimental ``event''.

The event $x$ was analyzed by assuming that the ``resonance''
$i_m$, where the maximum of the ratio $x_i/\varepsilon_i$ occurs,
is
a p$_{1/2}$ resonance and can serve as the starting point of the
analysis in the above sense. By help of the posterior distribution
$P(\xi|x)$, the shortest confidence interval $(\xi_<,\xi_>)$ was found such that
\begin{displaymath}
  \int_{\xi_<}^{\xi_>}d\eta P(\eta|x_1,\ldots,x_n)=0.68\,
\end{displaymath}
and it was recorded whether the true value $\hat{M}$ was inside
$(\mbox{max}(0,\sqrt{\xi_<^2-\varepsilon_1^2}),
\mbox{max}(0,\sqrt{\xi_>^2-\varepsilon_1^2}))$. 
This procedure was applied to $10^4$ vectors
$x$. On fig. \ref{fig:Bayes-Likelihood}, the symbols labeled
``Bayes'' give the relative frequency of ``success'', i.e the
probability to find $\hat{M}$ in  the above mentioned range
performed for different $r$ which controls the experimental
error. In this way, the curve on the figure was generated. (The
full and dashed lines are 4th order polynomials fit to the
points). Because of the lack of symmetry, the result is not
identical but only close to 0.68 (the dotted line).
However, in the limit of $r\rightarrow 0$ which means
$\varepsilon_i\rightarrow 0$ for all $i$, the criterion is obeyed
exactly~--- as it should be, since scale invariance is recovered in
this limit.

In comparison,  a ``likelihood'' analysis was performed.
This type of analysis is very popular. It amounts to the Bayesian
procedure with the prior distribution set constant. 
In the present
case, we had to cut off the posterior distribution for $M\gg\hat{M}$ in
order to normalize it very much as in \cite{Bowman:1990,Frankle:1992,Zhu:1992}. 
The figure shows, that the
likelihood method is inferior to even the approximate Bayesian
method. For good data, i.e. $r\rightarrow0$, the likelihood
method yields confidence intervals that are~--- in the light of
the criterion~--- too wide.
 For  data of marginal quality ($r\approx 1$), it yields no
reliable confidence intervals, because the frequency of successes is
considerably lower than the prescribed confidence.

\section*{Summary}
We have defined objective Bayesian statistics by supplementing
the Bayesian argument with the requirement that it should yield
``objectively correct'' confidence intervals. A theorem by
Stein~--- which can be found in the published mathematical
literature \cite{Stein:1965} but which seems to be unknown in
practice~--- shows that this requirement can be met provided that
the conditional probability $p(x|\xi)dx$ posses the symmetry
(\ref{invariance}) defined by a Lie group. By way of an example
we have extended objective Bayesian statistics to the common case
in which $p(x|\xi)dx$ does not have an exact symmetry. In the
example, a complex event $x=(x_1,\ldots,x_n)$ is broken down into
elementary ones $x_i$ among which there is at least one~--- say
$x_1$~--- whose conditional probability $p(x_1|\xi)dx$ possesses a
symmetry (\ref{invariance}). It is used to define the prior
$\mu(\xi)$. This has been termed approximate Bayesian
procedure. We have shown numerically that the AB procedure is
superior to the popular likelihood method as judged by the
objectivity of the deduced confidence intervals.

Note that the arguments presented here amount to a reconciliation of
the subjective and frequentist interpretations of probability.  The
Bayesian argument attributes a distribution to an object, i.e. the
hypothesis, which is given by Nature once and for all. This is
justified by interpreting probability distributions as a
representation of subjective knowledge on that object. The frequentist
interpretation insists that a probability distribution must be
verifiable~--- at least in a gedanken experiment~--- as a frequency
distribution that occurs in some stochastic process. We have described
a gedanken experiment to generate a distribution of Bayesian
confidence intervals from data that are conditioned by a fixed true
value of the hypothesis. The rate of success, i.e. of the true value
lying inside the confidence interval, turns out to be independent of
the true value, moreover the rate of success is equal to the
confidence prescribed in the Bayesian procedure~--- if the conditional
distribution possesses the symmetry (\ref{invariance}) and if the prior is
chosen to be the right Haar measure. Then the Bayesian
inference is found reasonable from a frequentist's point of view.
\nocite{Hartigan:1963}

\begin{thebibliography}{10}

\bibitem{Anderson:1992}
P.~W. Anderson, Physics Today {\bf Jan.},  {}9  (1992).

\bibitem{Dagostini:1995}
G. D'Agostini, Probability and Measurement. Uncertainity in Physics -- a
  {B}ayesian Primer, preprint {DESY} 95-242, {ISSN} 0418-9833 (1995).

\bibitem{Berger:1985}
J.~O. Berger, {\em Statistical Decision Theory and Bayesian Analysis}, {\em
  Springer Series in Statistics}, zweite ed. (Springer, New York, 1985).

\bibitem{Lee:1989}
P.~M. Lee, {\em Bayesian Statistics: An Introduction} (Oxford University Press,
  Oxford, 1989).

\bibitem{Press:1989}
J.~S. Press, {\em Bayesian Statistics: Principles, Models, and Applications},
  {\em Wiley Series in Probability and Mathematical Statistics} (Wiley, New
  York, 1989).

\bibitem{Howson:1993}
C. Howson and P. Urbach, {\em Scientific Reasoning: The Bayesian Approach}
  (Open Court, Chicago, 1993).

\bibitem{Bernado:1993}
J. Bernado and A. Smith, {\em Bayesian Theory}, {\em Wiley Series in
  Probability and Mathematical Statistics} (Wiley, Chichester, 1993).

\bibitem{Maximum}
See Conference Reports on ``Maximum-Entropy and Bayesian Methods'', Dordrecht,
  Reidel, annual published.

\bibitem{ISO:92}
{\em Guide to the expression of uncertainity in measurement}, International
  Organization for Standardization (ISO), Geneva, Switzerland, ISBN
  92-67-10188-9.

\bibitem{DIN:85}
DIN 1319 Teil 4: Grundbegriffe der Me{\ss}technik: Behandlung von
  Unsicherheiten bei der Auswertung von Messungen, Beuth Verlag Berlin, 1985,
  an English translation is available from the authors of
  \protect{\cite{Weise:1992}}.

\bibitem{Weise:1992}
K. Weise and W. W{\"o}ger, Meas. Sci. Technol. {\bf 3},  1  (1992).

\bibitem{Barlow:1989}
R.~J. Barlow, {\em Statistics} (Wiley, Chichester, 1989).

\bibitem{Welch:1963}
B.~L. Welch and H.~W. Peers, J. Royal Statist. Soc. B {\bf 25},  318  (1963).

\bibitem{Stein:1965}
C.~M. Stein,  in {\em Proc. Int. Research Seminar}, Statistical Laboratory,
  University of California, Berkley 1963, edited by J. Neyman and L.~L. Cam
  (Springer, New York, 1965), p.\ 217.

\bibitem{Villegas:1977b}
C. Villegas, J. Amer. Statist. Assoc. {\bf 72},  453  (1977).

\bibitem{Villegas:1981}
C. Villegas, Ann. Statist. {\bf 9},  768  (1981).

\bibitem{Villegas:1990}
C. Villegas, J. Amer. Statist. Assoc. {\bf 85},  1159  (1990).

\bibitem{Eadie:1971}
W. Eadie {\it et~al.}, {\em Statistical Methods in Experimental Physics} (North
  Holland, Amsterdam, 1971).

\bibitem{Bayes:1763}
T. Bayes, Phil. Trans. {\bf 53},  370  (1763), reprinted in Biometrica
  \textbf{45}, 296 (1958).

\bibitem{Villegas:1977a}
C. Villegas, J. Amer. Statist. Assoc. {\bf 72},  651  (1977).

\bibitem{Bowman:1990}
J.~D. Bowman {\it et~al.}, Physical Review Letters {\bf 65},  1192  (1990).

\bibitem{Frankle:1992}
C.~M. Frankle {\it et~al.}, Physical Review C {\bf 46},  778  (1992).

\bibitem{Zhu:1992}
X. Zhu {\it et~al.}, Physical Review C {\bf 46},  768  (1992).

\bibitem{Bowman:1995}
J.~D. Bowman {\it et~al.}, preprint (unpublished).

\bibitem{Hartigan:1963}
J. Hartigan, Ann. Math. Statist. {\bf 35},  836  (1964).

\end{thebibliography}

\end{document}